\documentclass[12pt]{article}
\usepackage{latexsym,kec}
\usepackage{graphicx}
\begin{document}
\pagestyle{empty}
\begin{flushright}
NMCPP/99-19\\
\end{flushright}
\begin{center}
{\bf\Large \boldmath{\(B\!-\!L\)} Neutrinos}\\
\vspace*{0.65cm}
{\bf Kevin Cahill\footnote{kevin@kevin.phys.unm.edu
\quad http://kevin.phys.unm.edu/\(\tilde{\ }\)kevin/}}\\
\vspace{.5cm}
New Mexico Center for Particle Physics\\
Department of Physics and Astronomy\\
University of New Mexico\\
Albuquerque, New Mexico 87131-1156\\
\vspace*{0.20cm}
\end{center}
\begin{quote}
\begin{center}
{\bf Abstract}
\end{center}
Neutrino masses and mixings are analyzed in terms
of left-handed fields and a
\(6\times6\) complex symmetric mass matrix \(\cM\)
whose singular values are the neutrino masses.
An angle \(\theta_\nu\)
characterizes the kind of the neutrinos,
with \(\theta_\nu=0\) for Dirac neutrinos
and \(\theta_\nu=\pi/2\) for Majorana neutrinos.
At \( \theta_\nu = 0 \)
baryon-minus-lepton number is conserved.
If \( \theta_\nu \approx 0 \),
the six neutrino masses coalesce
into three nearly degenerate pairs.
Thus the tiny mass differences
exhibited in the
solar and atmospheric neutrino experiments
are naturally explained by the approximate
conservation of \(B-L\). 
Neutrinos are nearly Dirac fermions.   
\par
This \(B-L\) model leads to these predictions:
neutrinos oscillate mainly between flavor eigenfields
and sterile eigenfields, and so the  
appearance of neutrinos
and antineutrinos is suppressed;
neutrinos may well be of cosmological importance;
in principle the disappearance of \(\nu_\tau\)
should be observable;
and \(0\nu\b\b\) decay is suppressed by
an extra factor of 
\(10^{-5}\)
and so will not be seen in the
Heidelberg/Moscow, IGEX, GENIUS, or CUORE
experiments.
\end{quote}
\vfill
\begin{flushleft}
\today\\
\end{flushleft}
\vfill\eject
\section*{}

\setcounter{equation}{0}
\pagestyle{plain}
\par
The current wisdom on neutrinos is that 
the seesaw mechanism~\cite{mgmrm}
forces their masses to be very small.
This paper presents 
a rather different explanation
of the experimental facts
based upon the approximate conservation
of baryon-minus-lepton number, \(B-L\):
If \(B-L\) is almost conserved,
then the six two-component neutrino fields
form three nearly Dirac neutrinos, and 
the six neutrino masses coalesce 
into three nearly degenerate pairs.
\par
If there are three right-handed neutrinos,
then there are six left-handed fields, the three left-handed flavor
eigenfields \(\nu_e, \nu_\mu\), and \(\nu_\tau\) and the
charge conjugates of the three right-handed neutrinos.
The neutrino mass matrix
is then a \(6\times6\) complex symmetric matrix \(\cM\)
which admits
a singular-value decomposition \(\cM = U M V^\dgr\)\@.
The singular values 
are the six neutrino masses \(m_j\),
and the unitary matrix \(V^\dgr\)
describes the neutrino mixings.
\par
An angle \(\theta_\nu\) is introduced
that describes the kind of the neutrinos. 
Dirac neutrinos have \(\theta_\nu = 0\), and
Majorana neutrinos have \(\theta_\nu = \pi/2\)\@.
If all Majorana mass terms vanish,
that is if \(\theta_\nu = 0\),
then the standard model conserves 
\(B-L\), which is a global \(U(1)\) symmetry.
It is therefore natural in the sense of 't Hooft~\cite{tHooft}
to assume that \(\theta_\nu \approx 0 \)
so that this symmetry is only slightly broken.
The neutrinos then are nearly Dirac fermions
and their masses coalesce into three pairs
of almost degenerate masses.
Thus the approximate conservation of \(B-L\)
explains the tiny mass differences
seen in the solar and atmospheric neutrino experiments
without requiring the
neutrino masses to be absurdly small.
If one sets \(\theta_\nu \simeq 0.003\),
suppresses inter-generational mixing,
and imposes a quark-like mass hierarchy,
then one may fit the essential features of the
solar, reactor, and atmospheric neutrino data
with otherwise random mass matrices \(\cM\) in the eV range.
Thus neutrinos easily can have masses
that saturate the cosmological bound of about 8 eV\@.
Moreover because neutrinos are almost Dirac fermions,
neutrinoless double-beta decay is suppressed by 
an extra factor
\(\sim \sin^2\theta_\nu \, \sin^2\phi_\nu \lsim 10^{-5}\),
where \(\phi_\nu\) is a second neutrino angle,
and is very slow,        
with lifetimes in excess of \(2\times10^{27}\) years.
\par
This \(B-L\) model of neutrino masses and mixings
leads to these predictions about future experiments:
The three flavor neutrinos
oscillate mainly into the conjugates of the right-handed fields,
which are sterile.  Thus all experiments that look
for the appearance of neutrinos will yield small or null signals,
like those of LSND and KARMEN. 
Secondly because neutrino masses are not required
to be nearly as small as the solar and atmospheric
mass differences might suggest,
neutrinos may well be an important part of hot dark matter.
Thirdly if a suitable experiment can be designed,
it should be possible to see the tau neutrino disappear.
Fourthly, the rate of neutrinoless double-beta decay
is suppressed by an extra factor
\(\sim \sin^2\theta_\nu \, \sin^2\phi_\nu \lsim 10^{-5}\)
and hence will not be seen in the
Heidelberg/Moscow, IGEX, GENIUS, or CUORE
experiments.

\section*{Masses and Mixings}
Because left- and right-handed fields transform differently
under Lorentz boosts, they cannot mix.
It is therefore convenient to write the
action exclusively in terms of two-component, left-handed
fields.  The two-component, left-handed neutrino
flavor eigenfields \( \nu_e, \nu_\mu, \nu_\tau\)
will be denoted \( \nu_i,\) for \(i = e, \mu, \tau\)\@.
The two-component, left-handed fields that are the
charge conjugates of the putative right-handed neutrino fields
\( n_{re}, n_{r\mu}, n_{r\tau}\) will be denoted
\( n_i = - i \s^2 \, n_{ri}^\dgr\) for \(i = e, \mu, \tau \),
where \( \s^2 \) is the second Pauli spin matrix\@. 
\par
The six left-handed neutrino fields 
\( \nu_i, n_i \) for \( i=1, 2, 3\)
can have three kinds of mass terms:
The fields \( \nu_i \) and \( n_j \)
can form the Dirac mass terms 
\(
i D_{ij} \nu_i \s^2 n_j - i D_{ij}^* n_j^\dgr \s^2 \nu_i^\dgr 
\); in a minimal extension of the standard model,
the complex numbers \(D_{ij}\) 
are proportional to the mean value
in the vacuum of the neutral component of
the Higgs field.
The fields \( n_i \) and \( n_j \) can form
the Majorana mass terms 
\(
i E_{ij} n_i \s^2 n_j - i E_{ij}^* n_j^\dgr \s^2 n_i^\dgr
\),
which break \(B-L\)\@.
Because these mass terms connect right-handed neutrino fields,
which are sterile,
they do not affect neutrinoless double-beta decay,
at least in leading order.
Within the standard model,
the complex numbers \( E_{ij} \)
are simply numbers;
in a more unified theory,
they might be proportional to the mean values
in the vacuum
of neutral components of Higgs bosons.
The fields \( \nu_i \) and \( \nu_j \)
can form the Majorana mass terms
\(
i F_{ij} \nu_i \s^2 \nu_j - i F_{ij}^* \nu_j^\dgr \s^2 \nu_i^\dgr
\), 
which break \(SU(2)_L\otimes U(1)_Y\) and \(B-L\)\@.
Because these mass terms connect left-handed neutrino fields,
they potentially drive neutrinoless double-beta decay.
In a minimal extension of the standard model,
the complex numbers \( F_{ij} \) might
be proportional to the mean values
in the vacuum of the neutral component
of a new Higgs triplet \( h_{ab} = h_{ba} \)\@.
\par
Since \( \s^2 \) is antisymmetric
and since any two fermion fields 
\( \chi \) and \( \psi \) anticommute,
it follows that 
\(
\chi \s^2 \psi = \psi \s^2 \chi\) and
\( \chi^\dgr \s^2 \psi^\dgr = \psi^\dgr  \s^2 \chi^\dgr \),
which implies that
the \(3 \times 3\) complex matrices \( E \) and \( F \)
are symmetric
\(
E^{\top} = E \quad \mathrm{and} \quad F^{\top} = F
\)
and that 
\( 
i D_{ij} n_j \s^2 \nu_i = i D_{ij} \nu_i \s^2  n_j.
\)
Thus if we introduce
the \( 6 \times 6 \) matrix 
\beq
\cM = \pmatrix{ F & D \cr
                D^\top & E\cr}
\label {cM}
\eeq
and the (transposed) six-vector 
\( N^\top = (\nu_e , \nu_\mu , \nu_\tau , n_e , n_\mu , n_\tau ) \)
of left-handed neutrino fields,
then we may gather the mass terms 
into the matrix expression
\beq
\frac{i}{2} N^\top \cM \s^2 N 
- \frac{i}{2} N^\dgr \cM^* \s^2 N^\dgr.
\label {the 6x6 mass term}
\eeq
\par
The complex symmetric
mass matrix \( \cM \) is not normal unless 
the positive hermitian matrix \( \cM \cM^\dgr \) is real
because 
\(
[ \cM , \cM^\dgr ] = 2 i \, \Im m \left( \cM \cM^\dgr \right)
\)\@.
When the mass matrix \(\cM \) is real, 
it may be diagonalized
by an orthogonal transformation.
In general \( \cM \) is neither real nor normal;
but like every matrix, 
it admits a singular-value decomposition~\cite{LAPACK}
\beq
\cM = U M V^\dgr
\label {SVD}
\eeq
in which the \(6\times6\) matrices \( U \) 
and \( V \) are both unitary and the
\(6\times6\) matrix \( M \) is 
diagonal, 
\(
M = \mathrm{diag}(m_1,m_2,m_3,m_4,m_5,m_6)
\),
with singular values \( m_j \ge 0 \),
which will turn out
to be the masses of the six neutrinos.
\par
The free, kinetic action density of 
a two-component left-handed spinor \( \psi \)
is \( i \psi^\dgr 
\left( \partial_0 - \vec \s \cdot \nabla \right)
\psi \)\@.
Thus by including the mass terms (\ref {the 6x6 mass term}),
one may write the free action density of
the six left-handed neutrino fields \( N \) as
\beq
\cL_0 = i N^\dgr 
\left( \partial_0 - \vec \s \cdot \nabla \right)
N
+ \frac{i}{2} N^\top \cM \s^2 N 
- \frac{i}{2} N^\dgr \cM^* \s^2 N^\dgr
\label {N0ad1}
\eeq
from which follow 
the equations of motion for \( N \) 
\beq
\left( \partial_0 - \vec \s \cdot \nabla \right)
N = \cM^* \s^2 N^\dgr
\label {dN=}
\eeq
and \(N^\dgr\)
\beq
\left( \partial_0 + \vec \s \cdot \nabla \right)
\s^2 N^\dgr = - \cM N.
\label {dN*=}
\eeq
Applying \( \left( \partial_0 + \vec \s \cdot \nabla \right) \)
to the field equation (\ref {dN=}) for \(N\)
and then using the field equation (\ref {dN*=}) for \(N^\dgr\),
we find that
\beq
\left( \Box - \cM^\dgr \cM \right) N = 0,
\label {TFE}
\eeq
in which we used the symmetry
of the matrix \( \cM \) to write \( \cM^* \)
as \( \cM^\dgr \)\@.
\par
The singular-value decomposition
\( \cM = UMV^\dgr \) allows us to express
this equation (\ref{TFE}) in the form
\beq
\left( \Box - M^2 \right) V^\dgr N = 0,
\label {TFE2}
\eeq 
which shows that 
the singular values \( m_i \) of the
mass matrix \( \cM \) are the neutrino masses
and that the eigenfield of mass \( m_j \) is
\beq
\nu_{m_j} = \sum_{i=1}^6 V^*_{ij} N_i.
\label {MEFs}
\eeq
The vector \( N_m \) of mass eigenfields is thus
\(
N_m = V^\dgr N
\),
and so
the flavor eigenfields \( N \) are given by
\(
N = V N_m
\)\@.
In particular, the three left-handed fields
\( \nu_i \) for \( i = e, \mu, \tau \) are
linear combinations of the six mass eigenfields,
\(
\nu_i = \sum_{j=1}^6 V_{ij} \nu_{m_j}
\)
and not simply linear combinations of three
mass eigenfields.  

\section*{Experimental Constraints}
The four LEP measurements of
the invisible partial width
of the \(Z\)
impose~\cite{LEP} upon the number of
light neutrino types the constraint
\( N_\nu = 2.984 \pm 0.008.
\)
The amplitude for the \(Z\) production of two
neutrinos \( \nu_{m_j} \) and \( \nu_{m_k} \) to lowest order
is 
\(
A(\nu_{m_j}, \nu_{m_k} ) \, \propto \sum_{i=1}^3 V^*_{ik} V_{ij}
\),
and therefore the x-section for that process is 
\(
\s(\nu_{m_j}, \nu_{m_k} ) \, \propto | \sum_{i=1}^3 V^*_{ik} V_{ij} |^2
\)\@.
The LEP measurement 
of the number \( N_\nu \) of light neutrino species
thus implies that the sum over the light-mass eigenfields is
\beq
\sum_{j,k \; \mathrm{light}}
| \sum_{i=1}^3 V^*_{ik} V_{ij} |^2
= 2.984 \pm 0.008. 
\label {constraint}
\eeq
\par
This constraint on the \(6\times6\) unitary matrix \(V\)
is quite well satisfied if all six neutrino masses
are light.  For in this all-light scenario, the sum is
\beq
\sum_{j,k=1}^6 \sum_{i=1}^3 \sum_{i'=1}^3
V^*_{ik} V_{ij}
V_{i'k} V^*_{i'j}
= \sum_{i=1}^3 \sum_{i'=1}^3 
\d_{ii'} \d_{ii'}
= \sum_{i=1}^3 1 = 3 
\simeq 2.984 \pm 0.008.
\label {all light}
\eeq
\par
If the Hubble constant in units
of 100 km/sec/Mpc is \(h \simeq 0.65 \), 
then the conservative upper bound
on the neutrino component of hot dark matter,
\( \Omega_\nu \lsim 0.2 \),
implies~\cite{Kayser,Kolb} that 
the sum of the masses of the light, stable 
two-component neutrinos that interact weakly is 
bounded by
\beq
\sum_{j \; \mathrm{light}} m_j \, \lsim \, 8 \, \mathrm{eV}.
\label {eubound}
\eeq
\par
The lowest-order amplitude for a neutrino \(\nu_i\)
to be produced by a charged lepton \(e_i\),
to propagate with energy \(E\)
a distance \(L\) as some light-mass eigenfield
of mass \( m_j \ll E \), and to produce a charged
lepton \( e_{i'} \) is 
\beq
A(\nu_i \to \nu_{i'}) \, \propto
\sum_{j\;\mathrm{light}}
V_{i'j} V^*_{ij}
e^{-\frac{im_j^2 L}{2E}}.
\label {nunuosc}
\eeq
The lowest-order amplitude for the 
anti-process, \( \bar \nu_i \to \bar \nu_{i'} \), 
involves the complex conjugate of the matrix \(V\)
\beq
A(\bar \nu_i \to \bar \nu_{i'})  \, \propto
\sum_{j\;\mathrm{light}}
V^*_{i'j} V_{ij}
e^{-\frac{im_j^2 L}{2E}}.
\label {bar(nunu)osc}
\eeq  
To lowest order      
the corresponding probabilities are 
\beq
P(\nu_i \to \nu_{i'}) 
\, \propto \!
\sum_{j,j'\,\mathrm{light}}
V_{i'j} V^*_{ij} V^*_{i'j'} V_{ij'} 
\exp{\left(\frac{i(m_{j'}^2 - m_j^2)L}{2E}\right)} 
\label {Pnunu'}
\eeq
and
\beq
P(\bar\nu_i \to \bar\nu_{i'}) \, \propto \!
\sum_{j,j'\,\mathrm{light}}
V_{i'j} V^*_{ij} V^*_{i'j'} V_{ij'} 
\exp{\left(\frac{-i(m_{j'}^2 - m_j^2)L}{2E}\right)}.
\label {Pbar(nunu')}
\eeq
If all six neutrinos are light,
then in the limit \(L/E \to 0\)
these sums are \(\d_{i i'}\)\@.
\par
If for simplicity we stretch the error bars
on the Chlorine experiment
and average over one year,
then the solar neutrino experiments,
especially Gallex and SAGE,
see a diminution of electron neutrinos
by a factor of about one-half:
\beq
P_{\mathrm{sol}}(\nu_e \to \nu_e ) \simeq \half,
\label {solexp}
\eeq 
which requires a pair of mass eigenstates
whose squared masses differ 
by at least \(\sim 10^{-10} \, \mathrm{eV}^2\)~\cite{Kayser}\@.
The reactor experiments, Palo Verde
and especially Chooz, imply that
these squared masses differ by less than
\( \sim 10^{-3} \, \mathrm{eV}^2\)~\cite{Kayser}\@.
\par
The atmospheric neutrino experiments,
Soudan II, Kamiokande III, IMB-3, and
especially SuperKamiokande,
see a diminution of muon neutrinos
and antineutrinos by about one-third:
\beq
P_{\mathrm{atm}}(\nu_\mu \to \nu_\mu ) \simeq \frac{2}{3},
\label {atmexp}
\eeq 
which requires a pair of mass eigenstates 
whose squared masses differ by 
\(10^{-3} \, \mathrm{eV}^2 \lsim |m_j^2 - m_k^2| 
\lsim 10^{-2} \, \mathrm{eV}^2\)~\cite{Kayser}\@.   

\section*{The \(B-L\) Model}
When the Majorana mass matrices \(E\) and \(F\)
are both zero, the action density (\ref {N0ad1})
is invariant under the \(U(1)\) transformation
\(
N' = e^{i\theta G} \, N
\)
in which the \(6\times6\) block-diagonal matrix 
\(
G = \mathrm{diag}(I,-I) \)
with \( I \) the \(3\times3\) identity matrix.
The kinetic part of (\ref {N0ad1}) is clearly
invariant under this transformation.
The mass terms are invariant only when
the anti-commutator
\beq
\{ \cM, G \} = 2 \pmatrix{ F & 0 \cr 0 & - E \cr} = 0
\label {0}
\eeq
vanishes.
\par
This \(U(1)\) symmetry is 
the restriction to the neutrino sector
of the symmetry generated by 
baryon-minus-lepton number, \(B-L\), 
which is exactly conserved in the standard model.
A minimally extended standard model
with right-handed neutrino fields \( n_{ri} \)
and a Dirac mass matrix \(D\) but with
no Majorana mass matrices,
\(E = F = 0\), also conserves \(B-L\)\@.
When \(B-L\) is exactly conserved, 
\emph{i.e.,} when \( D \ne 0 \) but \(E = F = 0\),
then the six neutrino masses \(m_j\)
collapse into three pairs of degenerate masses
because the left-handed and right-handed fields
that form a Dirac neutrino have the same mass.
\par
Suppose this symmetry is slightly broken by the Majorana
mass matrices \(E\) and \(F\).
Then for random mass matrices \( D \), \(E\), and \(F\),
the six neutrino masses \(m_j\) will form three
pairs of nearly degenerate masses
as long as the ratio
\beq
\sin^2\theta_\nu = \frac{\mathrm{Tr}( E^\dgr E + F^\dgr F )}
{\mathrm{Tr}( 2 D^\dgr D + E^\dgr E + F^\dgr F )}
\label {s}
\eeq
is small.   
For a generic mass matrix \(\cM\),
the parameter \(\sin^2\theta_\nu\) lies between the extremes
\(
0 \le \sin^2\theta_\nu \le 1
\)
and characterizes the kind of the neutrinos.
The parameter \(\sin^2\theta_\nu\) is zero for purely Dirac neutrinos
and unity for purely Majorana neutrinos.
\par
Let us now recall 't Hooft's definition~\cite{tHooft}
of naturalness: 
It is natural to assume that a parameter is small
if the theory becomes more symmetrical when the parameter vanishes.
In this sense it is natural to assume that 
the parameter \( \sin^2 \theta_\nu \)
is small because the minimally extended standard model
becomes more symmetrical, conserving \(B-L\),
when \( \sin^2 \theta_\nu = 0 \)\@.
\par
\begin{figure}
\centering
\includegraphics{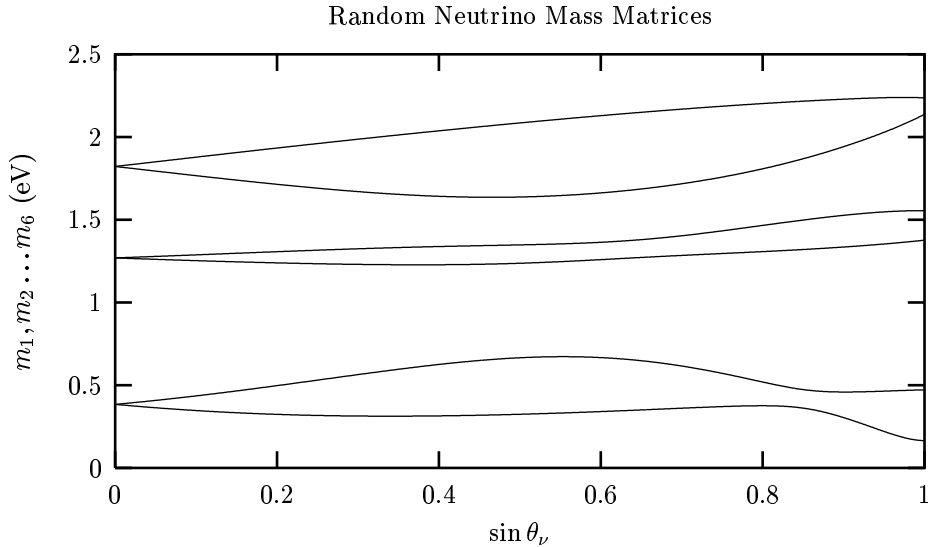}
\caption{The six neutrino masses are plotted against
the parameter \(\sin\theta_\nu\) for a set of 
random \(6\times6\) mass matrices.}
\label{fig:g1}
\end{figure}

In Fig.~\ref{fig:g1} the six neutrino masses \(m_j\)
are plotted for a set of mass matrices \(\cM\)
that differ only in the parameter \(\sin\theta_\nu\).
Apart from \(\sin\theta_\nu\),
every other parameter of the mass matrices \(\cM\) 
is a complex number \(z = x + i y \) in which
\(x\) and \(y\) were chosen
randomly and uniformly on the interval
\([ -1 \, \mathrm{eV}, 1 \, \mathrm{eV} ]\)\@.
It is clear in the figure that when \( \sin^2\theta_\nu \approx 0 \),
the six neutrino masses \(m_j\) coalesce into three nearly degenerate pairs.
Although the six masses of the neutrinos are in the eV range,
they form three pairs with very tiny mass differences
when \( \sin^2\theta_\nu \approx 0 \)\@.
\par
Thus the very small mass differences required by the
solar and atmospheric experiments are naturally
explained by the assumption that the symmetry
generated by \(B-L\)
is broken only slightly by the Majorana
mass matrices \(E\) and \(F\)\@. 
This same assumption implies that neutrinos are
very nearly Dirac fermions and hence explains
the very stringent upper limits~\cite{Kayser} on neutrinoless
double-beta decay.
Because the masses of the six neutrinos may lie
in the range of a few eV,
instead of being squashed down to
the meV range by the seesaw mechanism,
they may contribute to hot dark matter 
in a way that is 
cosmologically significant.
This \(B-L\) model with \( \sin^2\theta_\nu \approx 0 \)
is the converse of the seesaw mechanism. 
\par
\par
If \( \sin^2\theta_\nu = 0 \), then there are 
three purely Dirac neutrinos, 
and the mixing matrix \(V\) is block diagonal
\(
V = \mathrm{diag}( u^* ,  v )
\)
in which the \(3\times3\) unitary matrices \(u\) and \(v\)
occur in the singular-value decomposition of the
\(3\times3\) matrix \(D = u \, m \, v^\dgr \)\@.
If these three Dirac neutrinos are also light, 
then unitarity implies that 
the sum of the normalized probabilities is unity
\(
\sum_{i'=e}^\tau P(\nu_i \to \nu_{i'}) = 1
\)\@.
If \( \sin^2\theta_\nu = 1 \), then
this sum is also unity by unitarity 
because in this case the mixing matrix for the six
purely Majorana neutrinos is also block diagonal
\(
V = \mathrm{diag}\pmatrix{ v_F & v_E \cr}
\).
But if there are six light, nearly Dirac neutrinos,
then each neutrino flavor \(\nu_i\) will 
oscillate both into other neutrino flavor eigenfields
and into sterile neutrino eigenfields.
In this case this sum tends to be roughly a half
\(
\sum_{i'=e}^\tau P(\nu_i \to \nu_{i'}) \simeq \half
\) 
as long as \( \sin^2\theta_\nu \) is small but
not infinitesimal.
Because of this approximate, empirical sum rule~\cite{ndn} 
for \(i = e\) and \(\mu\),
the only way in which the probabilities
\( P_{\mathrm{sol}}(\nu_e \to \nu_i ) \)     
and \( P_{\mathrm{atm}}(\nu_\mu \to \nu_i ) \)
can fit the experimental results (\ref{solexp}) and (\ref{atmexp})
is if inter-generational mixing is suppressed
so that \( \nu_e \) oscillates into
\(n_e\) and so that \(\nu_\mu\)
oscillates into \(n_\mu\)\@.
In other words,
random mass matrices \( \cM \),
even with \( \sin\theta_\nu \approx 0 \),
produce probabilities \( P_{\mathrm{sol}}(\nu_e \to \nu_e ) \)
and \( P_{\mathrm{atm}}(\nu_\mu \to \nu_\mu ) \)
(suitably averaged respectively over the Earth's orbit and over 
the atmosphere) that are too small.
The probabilities \( P_{\mathrm{sol}}(\nu_e \to \nu_e ) \) 
and \( P_{\mathrm{atm}}(\nu_\mu \to \nu_\mu ) \)
do tend to cluster around \((\half,\frac{2}{3})\) as required
by the experiments when inter-generational mixing is 
severely repressed, that is if
the singly off-diagonal matrix elements of \(D, E,\) and \(F\)
are suppressed by 0.05 and the doubly off-diagonal matrix elements
by 0.0025\@. 
\par
\begin{figure}
\centering
\includegraphics{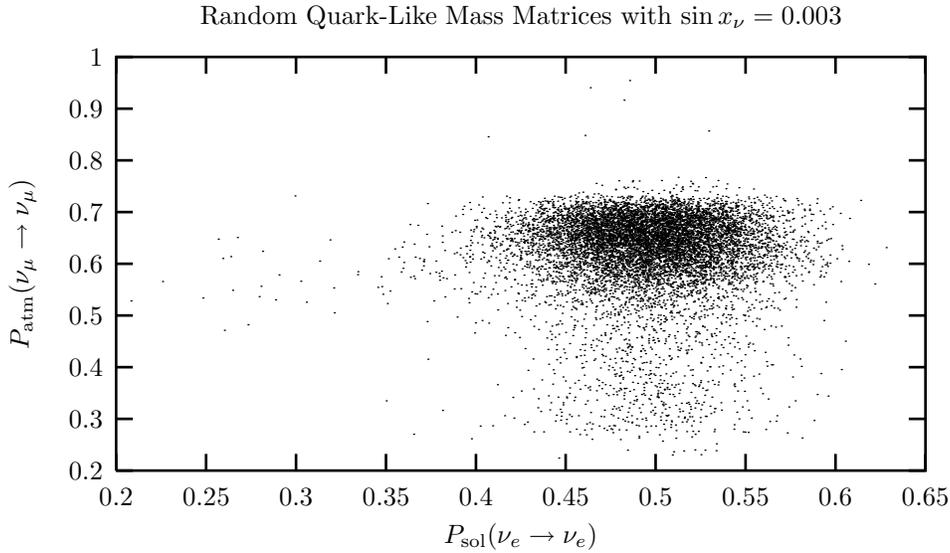}
\caption{The probabilities \( P_{\mathrm{sol}}(\nu_e \to \nu_e ) \)
and \( P_{\mathrm{atm}}(\nu_\mu \to \nu_\mu ) \)
for 10000 random mass matrices \(\cM\)
all with the parameter \(\sin\theta_\nu = 0.003\),
with inter-generational mixing suppressed, and
with a quark-like mass hierarchy.}
\label {fig:h5}
\end{figure}

It is possible to relax the factors that suppress
inter-generational mixing to 0.2 and 0.04
and improve the agreement with the
experimental constraints (\ref {solexp}) and (\ref {atmexp})
(while satisfying the CHOOZ constraint)
provided that one also requires 
that there be a quark-like mass hierarchy.
The points in Fig.~\ref{fig:h5}
were generated by random mass matrices
\( \cM \) with \( \sin\theta_\nu = 0.003 \)
by using CKM-suppression factors
of 0.2 and 0.04 
and by scaling the \(i,j\)-th elements
of the mass matrices \( E, F,\) and \(D\) 
by the factor \(f(i)*f(j)\) where
\( \vec f = ( 0.2, 1, 2 ) \)\@.
Thus the mass matrix \(\cM\)
has the \( \tau, \tau \)
elements that are larger than its
\( \mu, \mu \) elements and \( \mu, \mu \) elements
that in turn are larger than its \( e, e \) elements. 
The clustering  of 
the probabilities \( P_{\mathrm{sol}}(\nu_e \to \nu_e ) \)
and \( P_{\mathrm{atm}}(\nu_\mu \to \nu_\mu ) \) 
around \((\half,\frac{2}{3})\) in Fig.~\ref{fig:h5} shows that 
the experimental results (\ref {solexp}) and (\ref {atmexp}) 
are satisfied.
The vector \(\vec f\)
was tuned so as to nearly saturate 
the 
cosmological upper bound (\ref {eubound}) of about \( 8 \, \mathrm{eV}\)\@.
\par
In this scatter plot, 
every parameter of each of the 10000 matrices \(\cM\)
is a complex number \( z = x + i y \) with \(x\) and \(y\) chosen
randomly and uniformly from the interval \( [ - 1 \mathrm{eV},
1 \mathrm{eV} ] \)\@.
The solar neutrinos are taken to have an energy of 1 MeV,
and the probability (\ref {Pnunu'}) is averaged 
over one revolution of the Earth about the Sun.
The atmospheric neutrinos are taken to have an energy of 1 GeV,
and the probability (\ref {Pnunu'}) is averaged over the atmosphere
weighted by \(\sec \theta^*_Z\) in the notation
of Fisher \emph{et al.}~\cite{Kayser}\@.
The thousands of singular-value decompositions
were performed by the \textsc{lapack}~\cite{LAPACK} driver subroutine
\textsc{zgesvd}~\cite{lapack}\@.
\par
Neutrinoless double beta decay occurs when
a right-handed antineutrino emitted in one
decay \( n \to p + e^- + \bar \nu_e \)
is absorbed as a left-handed neutrino in the 
another decay \( \nu_e + n \to p + e^- \)\@. 
To lowest order these decays proceed via
the Majorana mass term 
\( - i F_{ee}^* \nu_e^\dgr \s^2 \nu_e^\dgr \)\@.
Let us introduce a second angle \(\phi_\nu\)
defined by
\beq
\sin^2\phi_\nu = \frac{\mathrm{Tr}( F^\dgr F )}
{\mathrm{Tr}( E^\dgr E + F^\dgr F )}.
\label {sf}
\eeq
We have seen that we may fit the experimental data
(\ref{solexp}) and (\ref{atmexp})
by assuming that \(\sin\theta_\nu \simeq 0.003\)
and by requiring the mass matrices 
\( E, F, \) and \(D\) to exhibit
quark-like mass hierarchies with little inter-generational mixing.
Under these conditions
the rate of \(0\nu\b\b\) decay is
limited by the factor
\beq
|F_{ee}|^2 \lsim \sin^2\theta_\nu \, \sin^2\phi_\nu \, m_{\nu_e}^2,
\label {0vbb}
\eeq
in which \( m_{\nu_e} \) is the heavier of the 
lightest two neutrino masses.
Thus the rate of \( 0\nu \b\b \) decay 
is suppressed by an extra factor 
\(\sim \sin^2\theta_\nu \sin^2\phi_\nu \lsim 10^{-5}\)
resulting in lifetimes \(T_{\half,0\nu\b\b} > 2\times10^{27}\,\)yr\@.
The \(B-L\) model therefore explains why
neutrinoless double-beta decay has not been seen
and predicts that the current and upcoming experiments
Heidelberg/Moscow, IGEX, GENIUS, and CUORE
will not see \( 0\nu \b\b \) decay.

\section*{Conclusions}
The standard model slightly extended to include
right-handed neutrino fields exactly conserves \(B-L\)
if all Majorana mass terms vanish.
It is therefore natural~\cite{tHooft} to assume that the Majorana mass terms 
are small compared to the Dirac mass terms.
A parameter \( \sin^2\theta_\nu \)
is introduced that characterizes the relative
importance of these two kinds of mass terms.
When this parameter is very small, then the neutrinos
are nearly Dirac and only slightly Majorana.
In this case the six neutrino masses \(m_j\)
coalesce into three pairs of nearly degenerate masses.
Thus the very tiny mass differences seen in the
solar and atmospheric neutrino experiments
are simply explained by the natural assumption that
\( \sin\theta_\nu \simeq 0.003 \) or equivalently that \(B-L\) 
is almost conserved.
In these experiments
the probabilities \( P_{\mathrm{sol}}(\nu_e \to \nu_e ) \)
and \( P_{\mathrm{atm}}(\nu_\mu \to \nu_\mu ) \) are respectively
approximately one half and two thirds.
One may fit these probabilities with random mass matrices 
in the eV range by requiring the neutrino mass matrices
\( E, F, \) and \(D\) to exhibit
quark-like mass hierarchies with little inter-generational mixing.
\par
This \(B-L\) model leads to these predictions: 
\begin{enumerate}
\item Because \( \sin^2\theta_\nu \approx 0 \) and
because inter-generational mixing is suppressed,
neutrinos oscillate mainly into sterile neutrinos 
of the same flavor and not into neutrinos of other flavors. 
Hence rates for the appearance of neutrinos, \(P(\nu_i \to \nu_{i'})\)
with \( i \ne i' \), are very low as shown by LSND and 
KARMEN\@.
\item The assumption that \( \sin^2\theta_\nu \)
is very small naturally explains the very small differences
of squared masses seen in the solar and atmospheric experiments
without requiring that the neutrino masses themselves be very small.
Thus the neutrinos may very well saturate 
the cosmological bound, \( \sum_j m_j \lsim 8 \, \mathrm{eV} \)\@.
In fact the masses associated with the points of Fig.~\ref{fig:h5}
do nearly saturate this bound.  Neutrinos thus may well be
an important part of hot dark matter.
\item The disappearance of \( \nu_\tau \) should \emph{in principle}
be observable.
\item In the \(B-L\) model, the rate of neutrinoless
double-beta decay
is suppressed by an extra factor 
\(\sim \sin^2\theta_\nu \, \sin^2\phi_\nu \lsim 10^{-5}\)
resulting in lifetimes greater than \(2\times10^{27}\,\)yr.
Thus the current and upcoming experiments
Heidelberg/Moscow, IGEX, GENIUS, and CUORE
will not see \( 0\nu \b\b \) decay.
\end{enumerate}

\section*{Acknowledgements}
I am grateful to H.~Georgi for
a discussion of neutrinoless double beta decay
and to B.~Bassalleck, J.~Demmel, B.~Dieterle,
M.~Gold, G.~Herling, D.~Karlen,
B.~Kayser, P.~Krastev, S.~McCready,
R.~Mohapatra, R.~Reeder, and G.~Stephenson for
other helpful conversations.

\end{document}